\documentclass[aps,twocolumn,superscriptaddress,pre,a4paper]{revtex4-1}

\fontencoding{T1}
\usepackage{epsfig}
\usepackage{amsfonts}
\usepackage{amsmath}
\usepackage{amssymb}
\usepackage{bm}
\usepackage{color}
\usepackage{graphicx}

\def\tst{t_\text{st}}

\begin{document}

\title{Self-propelled Vicsek particles at low speed and low density}

\author{M.\ Leticia {Rubio Puzzo}}

\affiliation{Instituto de F\'\i{}sica de L\'\i{}quidos y Sistemas Biol\'ogicos (IFLYSIB), 
CONICET y Universidad Nacional de La Plata, Calle 59 no.~789, B1900BTE La Plata, Argentina}
\affiliation{CCT CONICET La Plata, Consejo Nacional de Investigaciones Cient\'\i{}ficas y T\'ecnicas, Argentina}
\affiliation{Departamento de F\'isica, Facultad de Ciencias Exactas,
  Universidad Nacional de La Plata, Argentina}

\author{Andr\'es De Virgiliis}

\affiliation{Instituto de F\'\i{}sica de L\'\i{}quidos y Sistemas Biol\'ogicos (IFLYSIB), 
CONICET y Universidad Nacional de La Plata, Calle 59 no.~789, B1900BTE La Plata, Argentina}
\affiliation{CCT CONICET La Plata, Consejo Nacional de Investigaciones Cient\'\i{}ficas y T\'ecnicas, Argentina}
\affiliation{Departamento de Ciencias B\'asicas, Facultad de Ingenier\'\i{}a, Universidad Nacional de La Plata, Argentina}

\author{Tom\'as S. Grigera}

\affiliation{Instituto de F\'\i{}sica de L\'\i{}quidos y Sistemas Biol\'ogicos (IFLYSIB), 
CONICET y Universidad Nacional de La Plata, Calle 59 no.~789, B1900BTE La Plata, Argentina}
\affiliation{CCT CONICET La Plata, Consejo Nacional de Investigaciones Cient\'\i{}ficas y T\'ecnicas, Argentina}
\affiliation{Departamento de F\'isica, Facultad de Ciencias Exactas,
  Universidad Nacional de La Plata, Argentina}

\begin{abstract}

  We study through numerical simulation the Vicsek model for very low speeds and densities.  We consider scalar noise in 2-$d$ and 3-$d$, and vector noise in 3-$d$.  We focus on the behavior of the critical noise with density and speed, trying to clarify seemingly contradictory earlier results.  We find that, for scalar noise, the critical noise is a power law both in density and speed, but although we confirm the density exponent in 2-$d$, we find a speed exponent different from earlier reports (we consider lower speeds than previous studies).  On the other hand, for the vector noise case we find that the dependence of the critical noise cannot be separated as a product of power laws in speed and density.  Finally, we study the dependence of the relaxation time with speed and find the same power law in 2-$d$ and 3-$d$, with and exponent that depends on whether the noise is above or below the critical value.

\end{abstract}

\maketitle

\section{Introduction}

The Vicsek model (VM) \cite{VM} was proposed more than twenty years ago as a minimal model of flocking and swarming \cite{reviewVicsek}.  It has been since then widely studied \cite{ginelli}, and has established itself as sort of yardstick for flocking models.  Aside from its applications in understanding the microscopic mechanisms underlying swarming phenomena as observed in fish, birds, or mammals  \cite{sumpter_principles_2006}, it has attracted the attention of statistical physicists as a simple realization of a model of self-propelled particles (SPPs), i.e.\ out-of-equilibrium models where the speed of particles is maintained by a non-conservative source of energy \cite{marchetti_hydrodynamics_2013}.

In the VM each particle moves with a fixed speed $v_0$, and at each step the velocity is rotated so as to align with the average velocity of its neighbors (with some noise $\eta$ leading to non-perfect
alignment).  This aligning interaction leads to the development of order (flocking phase) at low noise and high number density ($\rho$), the order parameter (OP) being the system's average, or center-of-mass, velocity.  This is superficially similar to the order arising in lattice spin models such as Ising or Heisenberg, but a crucial feature of the VM is the coupling between density and order parameter \cite{toner_flocks_1998}: a density fluctuation that results in a local density higher that the critical one will result in a small cluster of ordered particles, but since ordered (i.e.\ velocity-aligned) particles travel together, these particles will tend to stay together, while ``capturing'' misaligned particles that by chance arrive in the neighborhood, thus enhancing density fluctuations.

This coupling of order parameter and density is largely responsible for the most salient features of the VM, namely \cite{ramaswamy_mechanics_2010} i) the existence of an order-disorder transition, controlled by density or noise, with the emergence of a phase with long-range order in the velocity, even in two dimensions, ii) the existence of propagating modes (density waves) in the orderedphase, and iii) a growth of the variance of the number of particles found in a given volume that grows faster than linearly in the number of particles (giant number fluctuations).

So in contrast to lattice spin models, the speed $v_0$ of the particles is more than simply a scale of measurement, because while the alignment interaction is independent of $v_0$, the displacement of
the particles in space is not, so that changing the speed alters the coupling of density and order parameter, and a change of $v_0$ cannot be compensated by a rescaling of time.  In fact the speed is a \emph{thermodynamic} parameter, since the critical values $\eta_c$ and $\rho_c$ of noise and number density at which the order-disorder transition occurs depend on $v_0$.  The aim of this article is to
study the thermodynamic and dynamic effects of variations in $v_0$ in the low density, low speed regime.

When $v_0=0$ (but keeping nonetheless a direction vector so that the interaction can be defined), the VM reduces in 3-$d$ to the classical Heisenberg model on a (random) graph (XY model in 2-$d$). For enough low density, most particles will be disconnected and the system will remain disordered for all values of the noise.  One thus expects $\eta_c\to0$ for $v_0\to0$ at low densities, but the exact dependence of $\eta_c$ with $\rho$ and $v_0$, as well as the dynamical effects of the reduction in speed, have not been thoroughly studied up to now.

The findings of published studies can be summarized as giving a power-law dependence of the critical noise on both speed and density,
\begin{equation}
  \label{eq:4}
  \eta_c\sim v_0^\sigma \rho^\kappa,
\end{equation}
although not all works study both variables simultaneously.  There are however differences in the reported values of the exponents, as well as in the theoretical arguments supporting them. Czir\'ok et al.\ \cite{czirok,czirok99} were the first to report a power law dependence of $\eta_c$, they found numerically $\kappa=0.25(5)$ in 1-$d$ and $\kappa=0.45(5)$ in 2-$d$.

Some time later, Chat\'e et al. \cite{chate} studied the phase diagram in the $(\eta,\rho,v_0)$ parameter space.  They argued that in the diluted limit ($\rho \ll 1/r_c^{d}$ where $r_c$ is the interaction radius) the critical value of the noise should behave as $\eta_c \sim v_0 \rho^{1/d}$, i.e.\ $\kappa=1/d$, $\sigma=1$.  The $\rho$ dependence was confirmed numerically in 2-$d$ and 3-$d$, but the linear $v_0$ dependence was only tested in 2-$d$ and at relatively high speeds.  This is compatible with the findings of refs.~\cite{czirok,czirok99} for 2-$d$, but not with
$\kappa =0.25$ for 1-$d$.  However, the 1-$d$ version of the model was defined in these reference with some modifications that maybe responsible for the disagreement.  Baglietto and Albano \cite{baglietto3} argued from numerical simulation (in 2-$d$) that $\eta_c$ tends to a finite limit when $v_0\to0$, however the analysis was done at $\rho=0.25$, and $v_0\geq 5 \times 10^{-3}$ (about two orders of magnitude above the speed values analysed in the present study).  More recently, Ginelli \cite{ginelli} revisited the issue and gave a modified argument (reviewed below in Sec.~\ref{sec:trans-crit-noise}), arguing instead that $\eta_c \sim \sqrt{\rho}$, which agrees 
with ref.~\cite{chate} only in 2-$d$.  However, the 3-$d$ case was not examined in this work.

In summary, though there seems to be agreement that $\eta_c \sim \sqrt{\rho}$ in 2-$d$, the speed dependence, as well as the density dependence in different dimensions, deserve further consideration.  It should also be mentioned that there are two variants of the VM in common use, which introduce the noise in different ways (scalar noise and vector noise, explained below).  The works quoted above use either one of the variants, and it is not clear whether the kind of noise has some influence on the differences found.

In the present work, we revisited the Vicsek Model in 2-$d$ and 3-$d$, paying special attention to the $\eta_c$ dependence with speed and density, in the slow and diluted limit. 

The manuscript is organized as follows: Sec.~\ref{sec:model-simul-deta} reviews the definition of the Vicsek Model and gives details of the simulations, results are presented in Sec.~\ref{sec:results-discussion},  and Sec.~\ref{sec:conclusions} states our conclusions.

\section{Model and simulation details}
\label{sec:model-simul-deta}

The Vicsek model consists of $N$ self-propelled particles endowed with a fixed speed $v_0$ and moving in $d$-dimensional space.  At each time step, positions $\mathbf{r}_i(t)$ and velocities $\mathbf{v}_i(t)$ are updated according to
\begin{align}
  \mathbf{v_i}(t+\Delta t) &= v_0 \mathcal{R}_\eta \left[ \sum_{j\in S_i}
                      \mathbf{v}_j(t) \right], \label{snoiseupd} \\
  \mathbf{r_i}(t+\Delta t) &= \mathbf{r}_i(t) + \Delta t
                             \mathbf{v}_i(t+\Delta t), \label{rFU}
\end{align}
where $S_i$ is a sphere of radius $r_c$ centered at $\mathbf{r}_i(t)$.
The operator $R_\eta$ normalizes its argument and rotates it
randomly within a spherical cone centered at it and spanning a solid
angle $\Omega_d\eta$, where $\Omega_d$ is the area of the unit sphere
in $d$ dimensions ($\Omega_2=2\pi$, $\Omega_3=4\pi$).

The order parameter, which measures the degree of flocking, is the normalized modulus of the average velocity
\cite{VM,reviewVicsek},
\begin{equation}
\varphi \equiv \frac{1}{N v_0} \left| \sum_{i=1}^{N} \mathbf{v}_i \right|.
\label{opVM}
\end{equation}
$\varphi\in[0,1]$, with $\varphi = O(1/\sqrt{N})\sim 0$ in the
disordered phase and $\varphi = O(1)$ in the the ordered phase.
We choose $\Delta t= r_c = 1$, so that the control parameters
are the noise amplitude $\eta$, the speed $v_0$ and the
number density $\rho = N/V$, where $V=L^d$ is the volume of the
(periodic) box.

The update rule for the positions, Eq.~\ref{rFU} is known as
\emph{forward update,} and was first used by Chat\'e \textsl{et al.\/}
\cite{chate}.  The original VM \cite{VM} used instead the so-called
\emph{backward update} rule, i.e.\
$\mathbf{r}_i(t+\Delta t)= \mathbf{r}_i(t) + \Delta t
\mathbf{v}_i(t)$).  It is generally agreed \cite{ginelli} that
choosing either prescription results in essentially the same behavior
for $t\to\infty$, although a small shift toward higher values of
$\eta_c$ has been reported for backward update \cite{baglietto2}.  We
have only used forward update in this work.

Equation~\ref{snoiseupd}, a $d$-dimensional generalization of the original direction update rule, uses what is known as \emph{scalar noise.} Scalar noise corresponds to a single source of noise in the alignment
(e.g.\ the local average velocity is measured exactly, but the adjustment of the direction is subject to noise).  An alternative rule which we also consider below is the so-called \emph{vector noise}, 
which corresponds to multiple sources of noise, e.g.\ in recording the velocity of each neighbor:
\begin{equation}
  \mathbf{v_i}(t+\Delta t) = v_0 \mathcal{N} \left[ \mathcal{N} \left( \sum_{j\in S_i}
      \mathbf{v}_j(t) \right) + {\pmb \xi} \right],
  \label{vnoiseupd}
\end{equation}
where $\mathcal{N}(\mathbf{v})=\mathbf{v}/\lvert \mathbf{v}\rvert$ and ${\pmb \xi}$ is a vector uniformly distributed on a sphere of radius $\eta$.

We performed standard Monte Carlo simulations of the VM in $d=2$ and $d=3$, using a simulation box of size $L^d$ with periodic boundary conditions.  In 2-$d$ we used densities $\rho=0.1$ and $\rho=1.0$ with $N=500$, $1000$ and $2000$ particles, while in 3-$d$, $N=1000$ and densities in the range $\rho=10^{-3}$ to $\rho=1$.  This corresponds to box sides $L=(N/\rho)^{1/d}$ in the range $L\simeq[22,140]$.  The range of speeds considered was from $v_0=10^{-1}$ to $v_0=10^{-5}$ in 2-$d$, and $v_0=1$ to $v_0=10^{-2}$ in 3-$d$.

Unless otherwise stated, simulations were started from a completely disordered initial condition, i.e.\ position and direction of motion ($\mathbf{r}_i(t=0)$ and $\mathbf{v}_i(t=0)$ ) chosen randomly.  In some cases we used a completely ordered initial state, where all particles are assigned the same velocity and distributed in a sphere of radius $2r_c$.

All results shown correspond to observables measured at the stationary state, which we have checked up to second order (i.e.\ for one- and two-time quantities).  We estimated the time needed to reach the stationary state in two different ways.  First, we recorded the time $\tst$ required for two systems with identical parameters, one starting from a completely disordered state ($\varphi(t=0)\sim0$) and another one starting from complete order ($\varphi(t>\tst)= 1$), to reach the same value $\varphi$.  
In addition, we estimated the correlation time from the (connected) time correlation function of the  order parameter,
\begin{equation}
C(t)= \big\langle \bigr(\varphi(t_0)- \langle\varphi\rangle)  (\varphi(t+t_0)- \langle\varphi\rangle \bigl)\big\rangle, \label{eq:tcorr}
\end{equation} 
where $\langle \ldots \rangle$ stands for an average over different simulation runs and time origins $t_0$.   The correlation time $\tau$ was then estimated from an exponential fit of the initial decay of $C(t)$.  Results for $\tst$ and $\tau$ are shown below (Sec.~\ref{sec:stat-state-corr}); essentially $\tst$ is about 4 to 5 times $\tau$.  In our measurements we have therefore discarded all data for $t<\tst$ and used time series of at least $10\tau$ ($100\tau$ in 2-$d$).  We have also checked that there are no aging effects in the time intervals studied, i.e.\ that $C(t)$ is independent of $t_0$.

\begin{figure}[!h]
\centering
\includegraphics[width=\columnwidth]{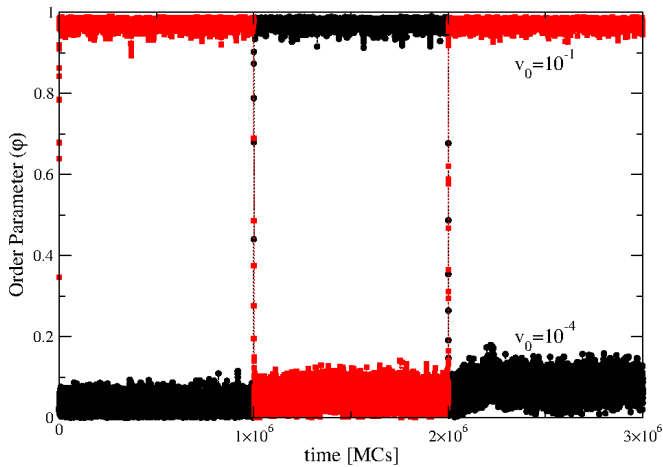}
\caption{Time evolution of the order parameter $\varphi$, for $\eta=0.01$, $N=10^3$ and $\rho=1.0$ by changing abruptly the speed $v_0$ between $10^{-1}$ to $v_0=10^{-4}$ (black circles), and from $10^{-4}$ to $v_0=10^{-1}$ (red squares).}
\label{figquench}
\end{figure}

As an additional precaution, for some values of $\rho$ and $v_0$ we have performed the following check:  after the OP had reached a stationary value in the ordered phase, we changed abruptly the speed to a value corresponding to the disordered phase, then back again to its original value.  A similar check but starting from a disordered state was also done (see Fig.~\ref{figquench}).  The absence of hysteresis confirms that the simulation times are such that we are investigating a stationary state independent of the initial conditions.


\begin{figure}[!h]
\centering
\includegraphics[width=\columnwidth]{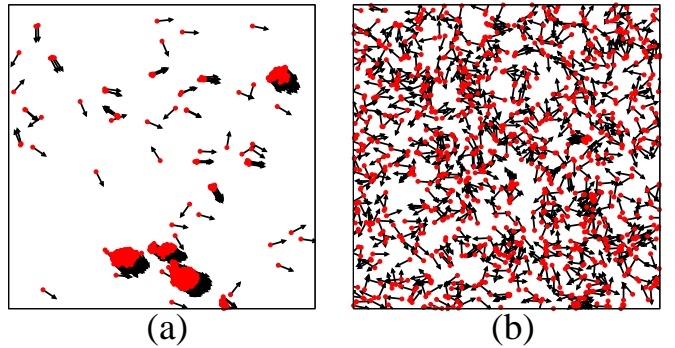}
\caption{Snapshots for $N=10^3$, $\rho=1.0$, $v_0=10^{-4}$, and $(a)$ $\eta=0.005<\eta_c(\rho,v_0)$; $(b)$ $\eta=0.015>\eta_c(\rho,v_0)$.}
\label{figsnp}
\end{figure}

\begin{figure}[!h]
\centering
\includegraphics[width=\columnwidth]{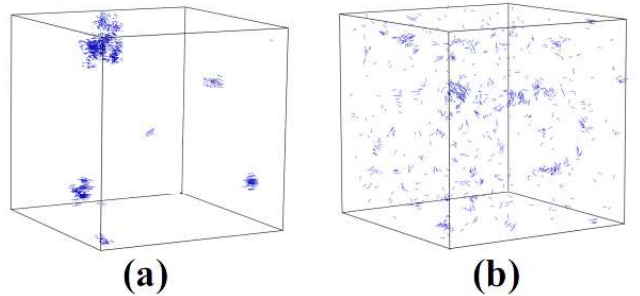}
\caption{Snapshots for $N=10^3$, $\rho=0.01$, $v_0=10^{-2}$, and $(a)$ $\eta=10^{-4}<\eta_c(\rho,v_0)$; $(b)$ $\eta=10^{-2}>\eta_c(\rho,v_0)$.}
\label{figsnp3d}
\end{figure}

Snapshots of typical ordered and disordered configurations are shown in Figs.~\ref{figsnp} (2-$d$) and \ref{figsnp3d} (3-$d$).

\section{Results and discussion}

\label{sec:results-discussion}

\subsection{Speed and density dependence of the critical noise}
\label{sec:trans-crit-noise}

To see why the critical noise depends on density and speed, one can give the following rough argument, for very dilute systems: the velocities of two particles close to each other (and isolated from the rest) will be roughly aligned, but the slight mismatch due to the noise will cause their relative position to move in a random walk until they become separated by a distance such that cease to interact and lose alignment.  The critical noise can be estimated as that at which the particles that have just ceased to interact encounter another particle and thus realign \cite{chate, ginelli}.  The relative distance behaves as $[\Delta \mathbf{r}(t)]^2 \sim \eta^2 t$, and the \emph{persistence length} (the distance traveled before the particles cease to interact) is
\begin{equation}
  \label{eq:2}
  l_P \sim v_0  / \eta^2.
\end{equation}
The distance traveled between collisions is the \emph{mean free path,} which scales as $l\sim 1/\rho$, so that equating the two lengths one gets
\begin{equation}
  \label{eq:3}
  \eta_c \sim  \sqrt{v_0 \rho}.
\end{equation}
This expression agrees with Eq.~9 of ref.~\cite{ginelli} (though the $v_0$ dependence is omitted in the reference), but not with the estimate $\eta_c\sim v_0 \rho^{1/d}$ of ref.~\cite{chate}, in which the noise amplitude instead of the variance was used to estimate $l_P$, and where the persistence length was compared to the average interparticle distance instead of the mean free path.

In this section we attempt to check Eq.~\ref{eq:4} and estimate the exponents, with emphasis on $\sigma$, which appears to have received less attention than $\kappa$.  In practice we study $\eta_c(v_0,\rho)$ at several densities.  We measure the average and variance of the order parameter, $\langle \varphi \rangle$ and $\text{Var}(\varphi)\equiv \langle \varphi^2\rangle - \langle\varphi\rangle^2$. The critical value of the noise, $\eta_c(v_0,\rho)$ was obtained as the point where $\text{Var}(\varphi)$ is maximum \footnote{At the sizes we consider, the transition appears second order, even if it might be first order in the thermodynamic limit.}.

\begin{figure}
  \centering
  \includegraphics[width=\columnwidth]{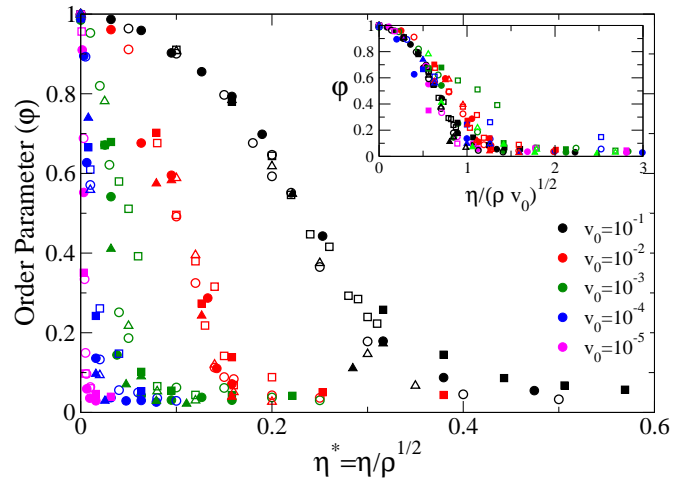}
  \caption{Order parameter as a function of the standard density-rescaled noise ($\eta^*\equiv \eta/\eta_c=\eta/\sqrt{\rho}$) for different $N$ (squares $500$, circles $N=1000$, and triangles $N=2000$), system densities $\rho$ (full symbols corresponds to density $\rho=0.1$ and open symbols to $\rho=1.0$) and velocities $v_0$ as indicated. Inset: The same data rescaled with a velocity-density-rescaled noise $=\eta^* v_0^{-0.45}$}
  \label{fig:OP2d}
\end{figure}

\begin{figure}
  \centering
  \includegraphics[width=\columnwidth]{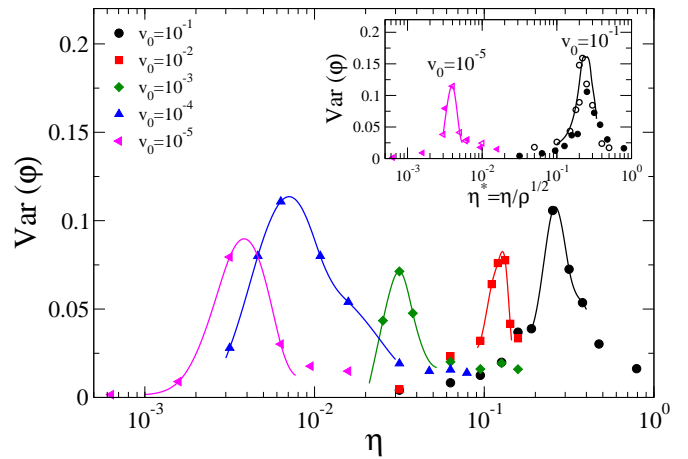}
  \caption{Log-linear plot of the fluctuations of the order parameter $\mathrm{Var}(\varphi)$ as a function of noise  $\eta$ for $N=1000$, $\rho=0.1$ and speed $v_0$ as indicated. Inset: The data collapsed with a density-rescaled noise $\eta^*=\eta/\sqrt{\rho}$, for two different speeds $v_0=10^{-5}$ and $v_0=10^{-1}$, and  densities $\rho=0.1$ (full symbols) and $\rho=1.0$ (open symbols).}
  \label{fig:chi2d}
\end{figure}

\begin{figure}
  \centering
    \includegraphics[width=\columnwidth]{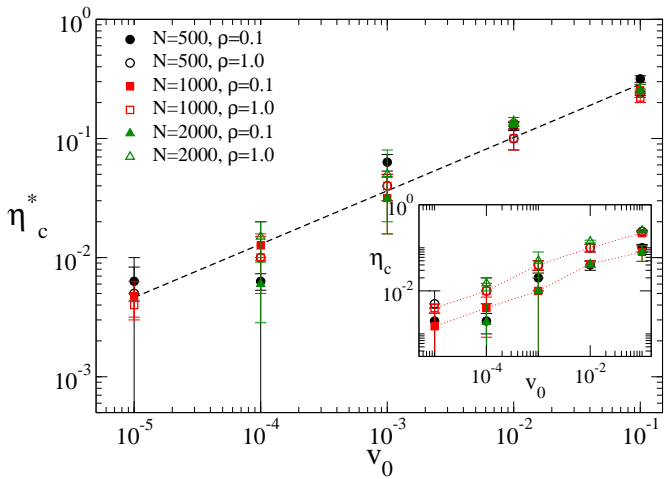}
    \caption{Density-rescaled-critical noise $\eta_c^*$ vs.\ speed $v_0$ for different densities and $N$, as indicated. Inset: Data before rescaling.}
    \label{fig:etac2d}
\end{figure}

\begin{figure}
  \centering
  \includegraphics[width=\columnwidth]{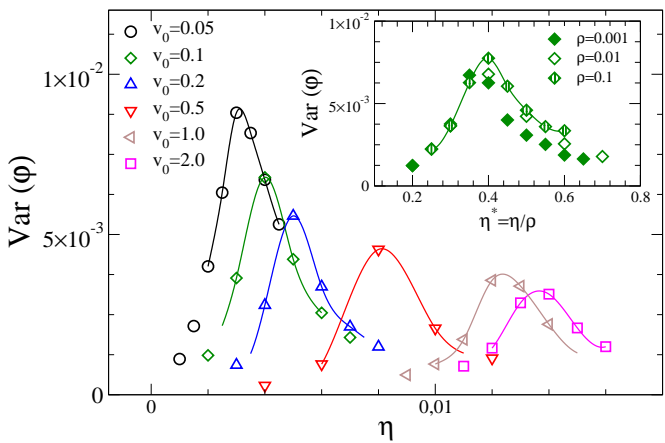}
  \caption{Fluctuations of the Order Parameter $Var(\phi)$ as a function of noise $\eta$, for $N=1000$, $\rho=0.01$ , and different $v_0$, as indicated. The value of $\eta$ that maximizes $Var(\phi)$ was taken as the critical noise $\eta_C$. Inset: Data collapsed $Var(\phi)$ as a function of noise $\eta*=\eta/\rho$ for $N=1000$, $v_0=0.1$, and three different densities as indicated.}
  \label{fig:chi3d}
\end{figure}

We consider first the scalar noise case in 2-$d$ and 3-$d$.  Figure~\ref{fig:OP2d} shows the order parameter and Fig.~\ref{fig:chi2d} its variance as a function of the noise (2-$d$), while  Fig.~\ref{fig:chi3d} shows the variance of the order parameter vs.\ noise in the 3-$d$ case.  Equation~\ref{eq:4} suggests defining a rescaled noise as
\begin{equation}
  \label{eq:6}
  \eta^*=\frac{\eta}{\rho^\kappa}.
\end{equation}
In 2-$d$ both $\langle\varphi\rangle$ and its variance scale reasonably well with $\eta^*$ (using $\kappa=1/2$)  for all the speeds considered (spanning four orders of magnitude).  This implies that $\eta_c\sim \sqrt{\rho}$ in 2-$d$, in agreement with earlier works \cite{czirok,chate, baglietto3, ginelli}.  Plotted as a function of $v_0$, the rescaled critical noise $\eta^*_c$ is also a power law $\sim v_0^\sigma$ (Fig.~\ref{fig:etac2d}), with a least-squares fit yielding $\sigma=0.45(2)$.  We have studied systems of $N=500$, 1000, and 2000 particles without finding significant differences.  Thus our 2-$d$ data are compatible with Eq.~\ref{eq:4}, with $\kappa=1/2$, $\sigma=0.45$.  This is in agreement with ref.~\cite{chate} for the $\kappa$ exponent, but not for $\sigma$, which these authors found close to 1.  However the speeds used in this article ranged from 0.05 to 0.5, while we have studied considerably smaller speeds, down to $v_0=10^{-5}$.

\begin{figure}
  \centering
  \includegraphics[width=\columnwidth]{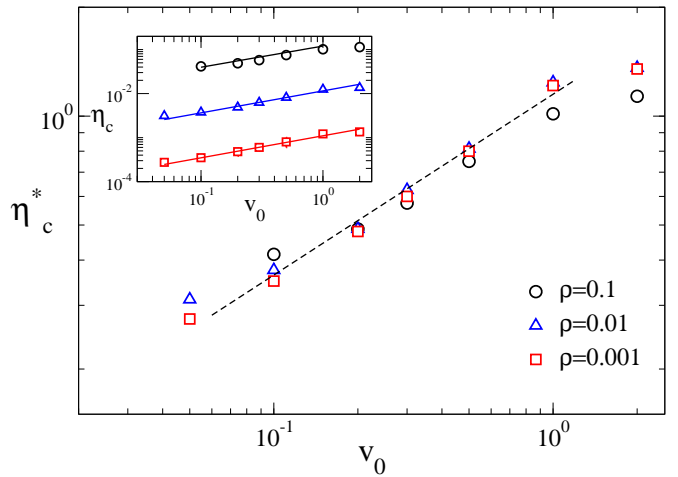}
  \caption{$\eta_C$ vs $v_0$ for N=1000 and different global-densities
    $\rho$, as indicated. Dashed lines has slope $1/2$ suggesting that
    $\eta_C \simeq \sqrt{v_0}$ in the 3D case.}
  \label{fig:etac3d}
\end{figure}

Turning to the 3-$d$, scalar noise case, we show $\text{Var}(\varphi)$ vs.\ $\eta$ in Fig.~\ref{fig:chi3d} and $\eta_c$ vs.\ $v_0$ in Fig.~\ref{fig:etac3d}.  The exponent $\sigma$ is obtained from a fit of $\eta_c$ vs.\ $v_0$ as $\sigma\approx 1/2$, i.e.\ very close to the 2-$d$ case.
We can collapse the $\eta_c$ vs.\ $v_0$ curves using the rescaled noise $\eta^*=\eta/\rho^\kappa$ (Fig.~\ref{fig:etac3d}), but with $\kappa=1$ instead of $1/2$.  In summary, we find that $\sigma$ is the same in 2-$d$ and 3-$d$ ($\sigma\approx 1/2$), while $\kappa$ is larger (1 rather that 1/2) in 3-$d$.

\begin{figure}
  \centering
  \includegraphics[width=\columnwidth]{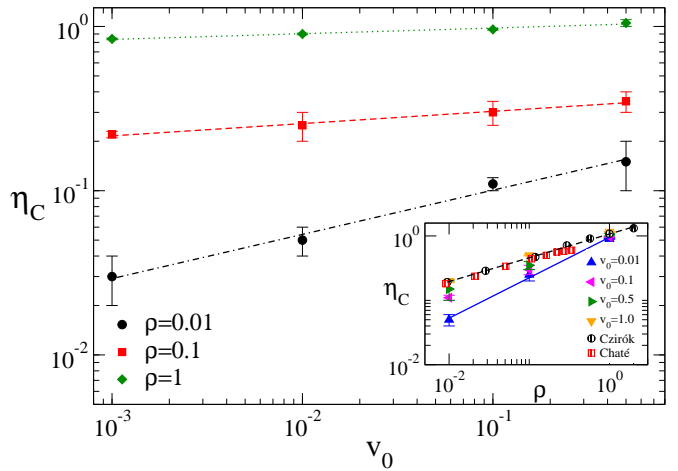}
  \caption{$\eta_C$ vs. $v_0$ for $N=1000$ and different global-densities
    $\rho$, as indicated, for the 3-dimensional Vicsek Model with vector noise. As can be seen, $\eta_C$ depends on $\rho$ as a power-law, but the exponent is clearly a non-trivial function of $\rho$. Inset: $\eta_c$ vs. $\rho$ for $N=1000$, and different $v_0$, as indicated. Plot includes points taken from refs. \cite{czirok3d} (Czir\'ok) and \cite{chate} (Chat\'e). Dashed line has slope $1/3$, in agreement with \cite{chate}; however, when speed $v_0$ slows down, the slope changes (full line has slope $0.63(4)$).}
  \label{fig:etac3dVN}
\end{figure}

The values of the exponents are rather different from those previously reported for the VM.  However, most previous studies used the vector variant of the VM, so we have also considered vector noise to investigate the effect of the noise rule on these exponents.  Figure \ref{fig:etac3dVN} shows $\eta_c$ vs.\ $v_0$ and $\eta_c$ vs.\ $\rho$ for the 3-$d$ Vicsek model with vector noise.  The figures also include points taken from refs.~\cite{chate} and~\cite{czirok3d}.  Our results reproduce the previously reported values of $\eta_c$ for $v_0=0.5$ \cite{chate}, but on going to lower speeds, we find strikingly that the slope of the logarithmic plots of the $\eta_c$ vs.\ $\rho$ curves depends on speed.  Similarly, the $\eta_c$ vs.\ $v_0$ slopes depend on density.  Thus it seems that the ansatz Eq.~\ref{eq:4} does not apply for the VM with vector noise.  At least down to $v_0=0.01$, the dependence of $\eta_c$ with noise and density \emph{is not separable,} i.e.\ cannot be written as product of a function of $\rho$ 
times a function of 
$v_0$.

\subsection{Stationary state and correlation time}
\label{sec:stat-state-corr}

\begin{figure}
  \centering
  \includegraphics[width=\columnwidth]{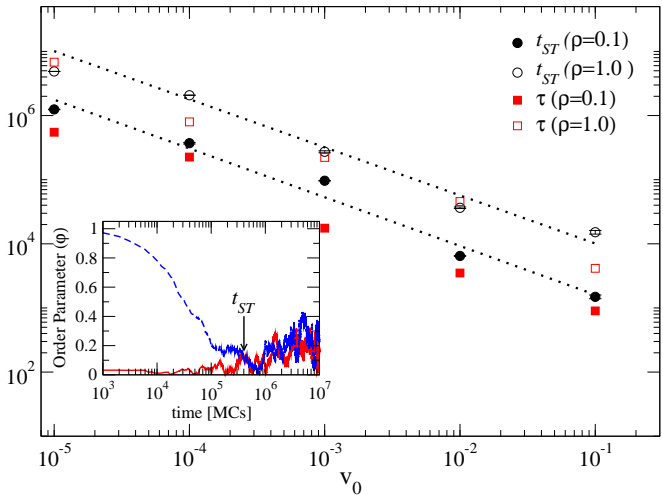}
  \caption{$\tst$ and $\tau$ vs.\ $v_0$ for $N=L^2=10^3$ and $\eta=\eta_c(v_0,\rho)$, and $\rho=0.1$ (open-symbols) and $\rho=1.0$ (full-symbols) and. Dotted-line has slope $-0.75$. Inset: Time evolution of the order parameter starting from disordered (full-line) and ordered (dashed-line) condition for $N=10^3$, $\rho=0.1$, and $v_0=10^{-4}$ at $\eta=\eta_c(v_0,\rho)$.}
  \label{fig:eqtime2}
\end{figure}

\begin{figure}
  \centering
  \includegraphics[width=\columnwidth]{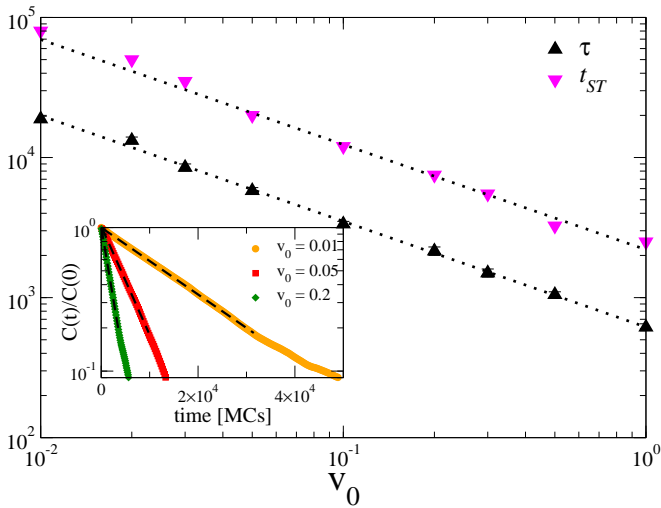}
\caption{$\tst$ and $\tau$ vs.\ $v_0$ for $N=L^3=10^3$ and $\eta=\eta_c(v_0,\rho)$, and $\rho=0.01$. Dotted-line has slope $-0.75$. Inset: Initial evolution of the Time Correlation function $C(t)/C(0)$ for $N=1000$, $\rho=0.01$, and different $v_0$, as indicated. Dashed line corresponds to the exponential decay fit ($C(t)\simeq e^{-t/\tau}$.}
  \label{fig:eqtime3}
\end{figure}

Turning back to the scalar noise case, our stationary-state checks allow us to investigate effects of speed on the dynamics of the VM by considering the $v_0$ dependence of $\tst$ and $\tau$, i.e. the time to reach a stationary state and the correlation time of the order parameter respectively (see Sec.~\ref{sec:model-simul-deta}).  The inset of Fig.~\ref{fig:eqtime2} shows how $\tst$ was determined from the convergence of the value of the order parameter in two systems starting from complete order and low density, and complete disorder and high density (see Sec.~\ref{sec:model-simul-deta}).  The case shown corresponds to the critical value of the noise.  The speed dependence of $\tst$ is  shown in Figs.~\ref{fig:eqtime2} and~\ref{fig:eqtime3} for 2-$d$ and 3-$d$ respectively.  The inset of Fig.~\ref{fig:eqtime3} shows a few instances of the time correlation function of the order parameter, Eq.~\ref{eq:tcorr}, from which $\tau$ is obtained by an exponential fit.  Figures ~\ref{fig:eqtime2} and~\ref{fig:eqtime3} show also $\tau$ vs.\ $v_0$ for 2 and 3 dimensions.

The result is that the dependence is well described by a power law for both quantities: $\tau$ and $\tst \propto v_0^\zeta$.  The exponent $\zeta$ depends on $\eta$ (but not on $\rho$ in the range we considered).  We find $\zeta\approx -0.75$ for $\eta=\eta_c$ and $\eta=0$, and $\zeta\approx -1$ for $\eta>\eta_c$.  Interestingly, the exponent is the same in 2-$d$ and 3-$d$.

\section{Conclusions}

\label{sec:conclusions}

In summary, we have studied the Vicsek model with scalar noise in two and three dimensions, and the VM with vector noise in three dimensions, at low densities and at lower speeds than in previous studies.  Our results support, for the scalar noise case in the diluted, low speed regime, the relation
\begin{equation}
  \eta_c\sim v_0^\sigma \rho^\kappa, \tag{\ref{eq:4}}
\end{equation}
but the exponents we find do not always agree with earlier studies, which have probed higher speeds than ours.  We find $\sigma \approx 1/2$ in both 2-$d$ and 3-$d$, at variance with previous reports.  For the other exponent we find $\kappa\approx 1/2$ in 2-$d$, as previously reported, and $\kappa\approx 1$ in 3-$d$.

Since most previous studies in 3-$d$ have used the vector noise variant, we have considered also the vector noise VM.  We have been able to reproduce earlier results, however, when exploring a wider range of speeds it becomes apparent that while at fixed density the curves look like a power law in speed (and vice-versa), the exponent depends on density, a signal that Eq.~\ref{eq:4} is not valid in this case, and that the dependence on speed and density is not separable.  These results show that the behavior of the diluted VM at low speeds is more complex that hitherto assumed.

In addition to the thermodynamic effect of speed, we have also investigated its effect on the dynamics, specifically the speed dependence of the relaxation time and the time to reach a stationary state independent of initial conditions.  We have found a power law for both quantities, with an exponent that depends on noise but not on density or space dimension ($\tau\sim v_0^{-1}$ for $\eta>\eta_c$ and $\tau\sim v_0^{-0.75}$ for $\eta=0$, $\eta=\eta_c$).  Thus the dynamics is slower for lower speeds, as one would expect naively given that information flow in $d<4$ is dominated by convective transport \cite{toner_flocks_1998}.  However, for fixed correlation length (e.g.\ at the critical point), one would guess naively $\tau v_0=\text{const}$, which is inconsistent with our findings.  In any case, the relaxation time for $v_0=0$ cannot be divergent, so either the relationship should break down at very low speeds, or the limit $v_0\to0$  is singular.

These results underline once more the complexities of the VM due to the coupling between density and order parameter, and call for a more detailed study at very low densities and speeds (particularly for the vector noise case), including (computationally expensive) considerations of finite-size effects.

\acknowledgements

We thank F.~Ginelli for discussions.  This work was supported by CONICET and UNLP (Argentina). 
Simulations were done on the cluster of \emph{Unidad de C\'alculo, IFLYSIB.}

\bibliography{newbib}

\end{document}